\documentclass[12pt,aps,prd]{revtex4}
\def\Aslash{{A\mkern-11mu/}}

\def\pslash{{p\mkern-8mu/}{\!}}

\def\prslash{{\partial\mkern-9mu/}}    
\def\qslash{{q\mkern-8mu/}{\!}}

\begin{document}

\title{Three dimensional noncommutative
bosonization}
\author{Rabin Banerjee}
\email{rabin@bose.res.in} \affiliation{S. N. Bose National Centre
for Basic Sciences, Block-JD, Sector-III, Salt Lake, Kolkata-700
098, India}
\author{Subir Ghosh}
\email{sghosh@isical.ac.in} \affiliation{Physics and Applied
Mathematics Unit, Indian Statistical Institute, 203 B. T. Road,
Kolkata-700 108, India}
\author{T. Shreecharan}
\email{shreet@bose.res.in} \affiliation{S. N. Bose National Centre
for Basic Sciences, Block-JD, Sector-III, Salt Lake, Kolkata-700
098, India}
\begin{abstract}
\begin{center}
{\small Abstract}
\end{center}
We consider the extension of the $2+1$-dimensional bosonization
process in Non-Commutative (NC) spacetime. We show that the large
mass limit of the effective action obtained by integrating out the
fermionic fields in NC spacetime leads to the NC Chern-Simons
action. The present result is valid to all orders in the
noncommutative parameter $\theta$. We also discuss how the NC
Yang-Mills action is induced in the next to leading order.
\end{abstract}

\maketitle

{\it Introduction}

Field theories in odd $(2+1)$ dimensional spacetime
\cite{djt,redlich} provide interesting features that are linked to
the presence of the Chern-Simons three form. It is known that even
if one does not include such a term from the beginning, it gets
induced as a result of quantum (one loop) effects \cite{redlich,first}. This, in
essence, is the phenomenon of bosonization in $(2+1)$ dimensions
\cite{fs,rb} since the Chern-Simons term appears in the effective
theory as one computes the fermion determinant in large mass
limit. This last point differentiates between bosonization in
$(1+1)$ dimensions \cite{col} where the fermion determinant is
exactly solvable and its $(2+1)$ dimensional counterpart, that can
only be determined locally as a power series in inverse fermion mass.

In this paper we discuss the issue of bosonization where the
underlying fermion model lives in a   noncommutative (NC)
spacetime, with the coordinates obeying
\begin{equation}
[x_{\mu},x_{\nu}]=i\theta_{\mu\nu}. \label{theta}
\end{equation}
The noncommutativity parameter $\theta_{\mu\nu}$ is a constant
antisymmetric object. In recent years NC  quantum field theories
have captured the interest of theoretical physics community ever
since their presence was established \cite{sw} in certain low
energy limits of open string theory in background field. In
particular NC theories in $(2+1)$ dimensions bear a special
interest since if one is only restricted to spatial NC with
$\theta_{0i}=0$, (as is generally the case), the minimal
spatial dimensionality has to be two. Infact the prototype of NC
theories is the Landau problem which deals with planar motion of
charged particles in a strong magnetic field.

A few years back a number of articles appeared \cite{sg,btc,wot}
that attempted to extend the well known duality, either in the Lagrangian\cite{dj} or Hamiltonian\cite{second} formulations, between
Maxwell-Chern-Simons theory and Self-Dual theory to their
respective NC generalizations. The issue was not conclusively
settled mainly for two reasons: firstly, computation of the NC
fermion determinant was not considered and the duality was studied
between the NC extended Maxwell-Chern-Simons theory and Self-Dual
theory. Secondly, for an explicit comparison one had to exploit
the Seiberg-Witten map \cite{sw} that connects NC fields to their
normal counterpart in the usual gauge theory. This makes the previous
analysis \cite{sg,btc,wot} perturbative in nature and only
$O(\theta )$ effects were taken into account since at higher
orders the Seiberg-Witten map itself is not unique.

In this perspective the present work becomes significant since we
have been able to rectify both the above mentioned drawbacks in a
single stroke. We have computed the $(2+1)$ dimensional NC fermion
determinant that is valid to all orders in the NC parameter
$\theta $. Obviously our NC extended  result is also valid in the
long wavelength (large fermion mass) limit. We have followed the
idea proposed in \cite{martin} where NC fermion effective actions
in $(3+1)$ dimensions were considered. It was demonstrated in
\cite{martin} that it is in fact enough to consider the existence
of an exact Seiberg-Witten map, valid to  all orders in $\theta $,
in a formal way and the explicit form of the map is not required.
As far as calculating the fermion effective action is considered,
the above procedure captures the NC effects in a non-perturbative
way. This makes the present analysis, {\it{i.e.}} NC bosonization,
exact to all orders in $\theta$.

To be more specific, we compute to all orders in $\theta$, the one
loop effective action in the large mass limit, that is obtained by
integrating out the fermionic matter sector in the NC gauge
theory. Due to the peculiarities of $(2+1)$ dimensions, the large
mass limit of the effective action turns out to be finite. In the
leading order, $\mathcal O(m^0)$, this is precisely the NC
Chern-Simons action. We also discuss, how in the next to leading
$\mathcal O(m^{-1})$ order, the NC Yang Mills term can be
obtained.

Our analysis shows that, in spite of the complexities of NC
spacetime, the one loop effective action (in the large mass limit)
yields exactly the NC extension of the commutative space result.
It might be recalled that the latter was also obtained in the
large mass limit\cite{fs,rb}.

{\it The effective action}

In NC spacetime, the action for the fermionic fields, in the
fundamental representation, in the presence of an external
non-abelian gauge potential, is given by
\begin{equation}
S = \int d^3x \left[\bar{\Psi} \star i \gamma^\mu D_\mu \Psi -m
\bar{\Psi} \star \Psi \right],
\end{equation}
where $(D_\mu)_{IJ} = \delta_{IJ} \partial_\mu - i (A_\mu)_{IJ}
\star$ and the star product is defined as,
\begin{equation}
(A \star B)(x)=\lim_{y\rightarrow
x}e^{\frac{i}{2}\theta^{\alpha\beta}\partial^x_{\alpha}\partial^y_{\beta}}A(x)B(y).
\label{star}
\end{equation}
The one-loop effective action is defined as
\begin{eqnarray}
i \Gamma[A] = {\rm ln} \frac{{\rm det}[\prslash + im - i \Aslash
\star]}{{\rm det}[\prslash + im]} = - \sum_{n=1}^{\infty}
\frac{1}{n} {\rm Tr} [(\prslash + im)^{-1} i \Aslash \star]^n
\label{4}
\end{eqnarray}
The operator $(\prslash + im)^{-1}$ gives the propagator for the
fermions
\begin{equation}
\langle y \mid (\prslash + im)^{-1} \mid x \rangle = \int
\frac{d^3 p}{(2 \pi)^3} \frac{i(\pslash + m)}{p^2 - m^2 + i
\epsilon} \, e^{-i p (y-x)}.
\end{equation}
which is identical to the commutative space expression. Hence,
\begin{equation} \label{eff1}
i \Gamma [A] \equiv \sum_{n=1}^\infty \int \!d^3 x_1 \cdots
\int\!d^3 x_n \, {\rm \textbf Tr}[A_{\mu_1}(x_1) \cdots
A_{\mu_n}(x_n)] \, \Gamma^{\mu_1 \cdots \mu_n} [x_1, \cdots, x_n],
\end{equation}
where,
\begin{eqnarray} \nonumber
\Gamma^{\mu_1 \cdots \mu_n}[x_1, \cdots, x_n] =
\frac{(-1)^{n+1}}{n} \int \prod_{i=1}^n \frac{d^3 p_i}{(2\pi)^3}
\,(2\pi)^3 \delta(\sum_i
p_i) \, e^{i \sum_{i=1}^n p_i x_i} \, e^{-i/2 \sum_{i<j}^n p_i \times p_j} \\
\int \frac{d^3 q}{(2\pi)^3}\,\frac{{\rm tr} [(\qslash +
\pslash_{\,\,1} + m) \, \gamma^{\mu_1} \, (\qslash + m) \,
\gamma^{\mu_2} \, (\qslash - \pslash_{\,\,2} + m) \,
\gamma^{\mu_3} \cdots (\qslash - \sum_{i=2}^{n-1} \pslash_{\, \,i}
+ m)\gamma^{\mu_n}]}{[(q + p_{\,1})^2 - m^2]\, [q^2 - m^2]\, [(q -
p_{\,2})^2 - m^2] \cdots [(q - \sum_{i=2}^{n-1}p_{\,i})^2-m^2]}.
\label{7}
\end{eqnarray}
Note that the effect of the star product in (\ref{4}) is
manifested in (\ref{7}) through the exponential phase factor with,
\begin{eqnarray}
p_i\times p_j=\theta^{\alpha\beta}p_{(i)\alpha}p_{(j)\beta}
\end{eqnarray}
The consistency of this formalism in NC space time and the fact
that there is no UV/IR mixing can be justified by exploiting the
Seiberg-Witten map that connects the NC field variables to their
commutative counterpart and vice-versa \cite{martin}. As we have
emphasized before, only a formal definition of the exact
Seiberg-Witten map is needed in this process.

The above two equations (\ref{eff1},\ref{7}) form the basis of our
calculations. Furthermore, for the sake of notational simplicity,
we have not written down the $i \epsilon$ terms in the propagators
and it is to be understood that they exist.

We now explicitly compute the effective action in the long
wavelength (i.e. large $m$) limit. This is done to the leading
$\mathcal O(m^0)$ and next to leading $\mathcal O(m^{-1})$ orders.
In the leading order there are only two contributions that arise
from the two-$A$ and three-$A$ terms in (\ref{eff1}). The other
pieces drop out due to the peculiarities of three dimensional
spacetime and the trace properties of gamma matrices. In the next
to leading order $\mathcal O(m^{-1})$, apart from the
contributions coming from the two-$A$ and three-$A$ terms, there
is also a piece from the four-$A$ term.
%
%
Equation (\ref{eff1}) is now explicitly evaluated order by order
in the background fields. The term with one gauge field vanishes
because of the tracelessness of the group matrices. Next, the
effect of the two-A term is considered.

{\it Contribution from two A terms}

First, we compute the odd parity contribution. The relevant
expression is obtained from (\ref{eff1}),
\begin{equation} \label{2aeff1}
i \Gamma [AA] = \int d^3 x_1 \, d^3 x_2 \, {\rm \textbf
Tr}[A_{\mu_1}(x_1) A_{\mu_2}(x_2)] \, \Gamma^{\mu_1 \mu_2} [x_1,
x_2];
\end{equation}
with
\begin{equation} \label{2aeff2}
\Gamma^{\mu_1 \mu_2}[x_1, x_2] = - \frac{1}{2} \int \frac{d^3
p_1}{(2\pi)^3} \frac{d^3 p_2}{(2\pi)^3} \, (2\pi)^3 \delta(p_1 +
p_2) \, e^{i (p_1 \cdot x_1 + p_2 \cdot x_2)} \, e^{-i(p_1 \times
p_2)/2} \, I^{\mu_1 \mu_2}.
\end{equation}
where,
\begin{equation}
I^{\mu_1 \mu_2} = \int \frac{d^3 q}{(2\pi)^3}\,\frac{{\rm tr}
[(\qslash + \pslash_{\,\,1} + m) \, \gamma^{\mu_1} \, (\qslash +
m) \, \gamma^{\mu_2}]}{[(q + p_{\,1})^2 - m^2]\, [q^2 - m^2]}.
\end{equation}
In what follows we will concentrate on the $q$ integral and its
explicit evaluation. The trace can be performed using the
identities provided in appendix A. We thus get
\begin{equation}
I^{\mu_1 \mu_2} = 2 \int \frac{d^3 q}{(2\pi)^3} \frac{\left[2
q^{\mu_1} q^{\mu_2} + p_1^{\mu_1} q^{\mu_2} + q^{\mu_1}
p_1^{\mu_2} + g^{\mu_1 \mu_2}(m^2 - q \cdot p_1 -q^2) + m p_{1
\alpha} \epsilon^{\alpha \mu_1 \mu_2} \right]}{[(q + p_{\,1})^2 -
m^2]\, [q^2 - m^2]} \label{11}
\end{equation}
Here we take the leading term in the large mass limit. In this
limit the $g^{\mu_1 \mu_2}$ term and the last term contribute.
Using the integral Eq. (\ref{2aint}) in the appendix, the odd
parity contribution is given by,
\begin{equation} \label{2a}
I^{\mu_1 \mu_2} = \frac{m}{|m|} \frac{i p_{1 \,\alpha}
\epsilon^{\mu_1 \mu_2 \alpha}}{4 \pi}
\end{equation}
Substituting equations (\ref{2a}) and (\ref{2aeff2}) in Eq.
(\ref{2aeff1}) we get
\begin{equation}
i \Gamma [AA] = - \frac{m}{|m|} \frac{\epsilon^{\mu_1 \mu_2
\mu_3}}{8 \pi}  \int d^3 x_1 \, d^3 x_2 \, {\rm \textbf
Tr}[A_{\mu_1}(x_1) A_{\mu_2}(x_2)] \partial^{x_1}_{\mu_3}
\delta(x_1-x_2).
\end{equation}
In deriving the above result we have performed both the $p$
integrations in (\ref{2aeff2}) that has yielded the delta
function. Performing an integration by parts and neglecting the
surface terms we get the final form,
\begin{equation} \label{2aresult}
i \Gamma [AA] = + \frac{m}{|m|} \frac{\epsilon^{\mu\nu\lambda}}{8
\pi}  \int d^3 x \, {\rm \textbf Tr}[A_{\mu}(x) \partial_{\nu}
A_{\lambda}(x)]
\end{equation}
It should be mentioned that the effect of the phase factor in
(\ref{7}) trivializes so that no star product occurs in
(\ref{2aresult}). This is expected because of the special property
\begin{equation}
\int d^3 x (A \star B)(x)=\int d^3 x (AB)(x)
\end{equation}
Observe that (\ref{2aresult}) is precisely the two-A part of the
NC Chern-Simons action, which is of $\mathcal O(m^0)$.

We now consider the normal parity part in (\ref{11}). This is
essentially the next to leading order $\mathcal O(m^{-1})$
contribution.

Using the integrals given in appendix B, one obtains,
\begin{eqnarray} \nonumber
I^{\mu_1 \mu_2} & = & - \frac{i}{16 \pi} \frac{1}{|m|}
\left[p_1^{\mu_1} p_1^{\mu_2} - m^2 g^{\mu_1 \mu_2} \right].
\end{eqnarray}
Putting the above result in Eq. (\ref{2aeff2}), we obtain,
\begin{equation}
\Gamma^{\mu_1 \mu_2}[x_1, x_2] = +  \frac{i}{32 \pi} \frac{1}{|m|}
\int \frac{d^3 p_1}{(2\pi)^3}\, e^{ip_1 \cdot( x_1 -
x_2)}\left[p_1^{\mu_1} p_1^{\mu_2} - m^2 g^{\mu_1 \mu_2} \right];
\end{equation}
which in turn is utilized in Eq. (\ref{2aeff1}) to yield,
\begin{equation}
i \Gamma [AA] =  \frac{i}{32 \pi} \frac{1}{|m|} \int d^3 x_1 \,
d^3 x_2 \, {\rm \textbf Tr}[A_{\mu_1}(x_1) A_{\mu_2}(x_2)] \int
\frac{d^3 p_1}{(2\pi)^3}\, e^{ip_1 \cdot( x_1 -
x_2)}\left[p_1^{\mu_1} p_1^{\mu_2} - m^2 g^{\mu_1 \mu_2} \right].
\end{equation}
Writing the momenta as derivatives, performing the $p_1$ integral
to yield a delta function, and finally doing an integration by
parts, the effective action simplifies to,
\begin{eqnarray} \nonumber
i \Gamma [AA] =  - \frac{i}{32 \pi} \frac{1}{|m|} \int d^3 x_1 \,
d^3 x_2 \, {\rm \textbf
Tr}\left[(\partial^{\mu_1}_{x_1} A_{\mu_1}(x_1)) (\partial^{\mu_1}_{x_2} A_{\mu_2}(x_2)) \right. \\
\left. - g^{\mu_1 \mu_2} (\partial^{\alpha}_{x_1} A_{\mu_1}(x_1))
(\partial_{\alpha}^{x_2} A_{\mu_2}(x_2)) \right] \delta( x_1 -
x_2), \\ \nonumber =  - \frac{i}{32 \pi} \frac{1}{|m|} \int d^3 x
\, {\rm \textbf
Tr}\left[(\partial^{\mu} A_{\mu}) (\partial^{\nu} A_{\nu}) -  g^{\mu \nu} (\partial^{\alpha} A_{\mu}) (\partial^{\alpha} A_{\nu}) \right], \\
=  \frac{i}{64 \pi} \frac{1}{|m|} \int d^3 x \, {\rm \textbf
Tr}\left[2 (\partial^{\mu} A_{\nu}) (\partial_{\mu} A^{\nu}) -  2
(\partial^{\mu} A_{\nu}) (\partial^{\nu} A_{\mu}) \right].
\label{28}
\end{eqnarray}
The above is the two-A term of the NC Yang-Mills action.

{\it Contribution from three-A terms}

The contribution of the three-A term in the effective action is
obtained from (\ref{eff1}),
\begin{equation} \label{3aeff1}
i \Gamma [AAA] = \int d^3 x_1 \, d^3 x_2 \, d^3 x_3 \, {\rm
\textbf Tr}[A_{\mu_1}(x_1) A_{\mu_2}(x_2) A_{\mu_3}(x_3)] \,
\Gamma^{\mu_1 \mu_2 \mu_3} [x_1, x_2, x_3],
\end{equation}
where
\begin{eqnarray} \nonumber
\Gamma^{\mu_1 \mu_2 \mu_3}[x_1, x_2, x_3] = \frac{1}{3} \int
\frac{d^3 p_1}{(2\pi)^3} \frac{d^3 p_2}{(2\pi)^3} \frac{d^3
p_3}{(2\pi)^3} \, (2\pi)^3 \delta(p_1 + p_2 + p_3)\\
\label{3adef}e^{i (p_1 \cdot x_1 + p_2 \cdot x_2 + p_3 \cdot x_3)}
\, e^{-i(p_1 \times p_2 + p_1 \times p_3 + p_2 \times p_3 )/2} \,
I^{\mu_1 \mu_2 \mu_3}.
\end{eqnarray}
In the present case
\begin{equation}
I^{\mu_1 \mu_2 \mu_3} = \int \frac{d^3 q}{(2\pi)^3}\,\frac{{\rm
tr} [(\qslash + \pslash_{\,\,1} + m) \, \gamma^{\mu_1} \, (\qslash
+ m) \, \gamma^{\mu_2} \, (\qslash - \pslash_{\,\,2} + m) \,
\gamma^{\mu_3}]}{[(q + p_{\,1})^2 - m^2]\, [q^2 - m^2]\, [(q -
p_{\,2})^2 - m^2]} \label{31}
\end{equation}
As before, we first compute the odd parity terms. These are
basically the leading order $\mathcal O(m^0)$ pieces. The
integrand is simplified as,
\begin{eqnarray} \nonumber
2im \, \left[\epsilon^{\alpha \mu_1 \mu_3} \left(q^{\mu_2}
q_\alpha + q^{\mu_2} p_{1 \alpha} - p^{\mu_2}_2 p_{1 \alpha}
\right) + \epsilon^{\alpha \mu_2 \mu_3} \left(q^{\mu_1} p_{2
\alpha} + p_1^{\mu_1} p_{2 \alpha} - q^{\mu_1} q_{\alpha} \right)
\right.
\\
- \left. \epsilon^{\alpha \mu_1 \mu_2} \left(p_2^{\mu_3} p_{1
\alpha} + q^{\mu_3} q_{\alpha} \right) + m^2 \epsilon^{\mu_1 \mu_2
\mu_3} \right]. \label{33}
\end{eqnarray}
The terms we are interested in are the ones independent of the
$p$'s and the last term of the above expression. Using (\ref{33})
and the integrals provided in equations (\ref{3aint1}) and
(\ref{3aint2}), we obtain from (\ref{31}),
\begin{equation} \label{3ares}
I^{\mu \nu \lambda} = + \frac{m}{|m|} \frac{\epsilon^{\mu \nu
\lambda}}{4 \pi}.
\end{equation}
Substituting equations (\ref{3ares}) and (\ref{3adef}) in Eq.
(\ref{3aeff1}), and using the definition (\ref{star}) of the star
product on a chain of fields,
\begin{eqnarray} \nonumber
\int dx \left[\mathcal{O}_1(x) \cdots \star \mathcal{O}_n(x)
\right] & = & \int \left[\prod_{i=1}^n dx_i\right]
\left[\prod_{j=1}^n \frac{dk_j}{(2 \pi)^D} \right] \exp\left(i
k^\mu_i \cdot x_\mu^i\right) \exp\left[- \frac{i}{2}
\sum_{i<j=1}^n k_\mu^i k_\nu^j \Theta^{\mu \nu}\right] \\ && (2
\pi)^D \delta(k_1 + k_2 + \cdots + k_n).
\end{eqnarray}
we obtain the final result,
\begin{equation} \label{3aresult}
i \Gamma [AAA] = + \frac{m}{|m|} \frac{2}{3} \frac{\epsilon^{\mu
\nu\lambda}}{8 \pi} \int d^3 x \,{\rm \textbf Tr}[A_{\mu}(x) \star
A_{\nu}(x) \star A_{\lambda}(x)].
\end{equation}
where the star product now appears explicitly. Combining equations
(\ref{2aresult}) and (\ref{3aresult}) we get the complete NC
Chern-Simons action.

In a similar way it is possible to compute the normal parity
contribution associated with the three-A term. In the order
$\mathcal O(m^{-1})$ expansion this yields the three-A piece in
the NC Yang-Mills action. Furthermore, there is an analogous
$\mathcal O(m^{-1})$ contribution from the four-A term in
(\ref{eff1}). Taking the two-A piece explicitly computed in
(\ref{28}), one obtains the complete structure of the NC
Yang-Mills action.

{\it Discussions} In the present work we have studied the
noncommutative extension of the bosonization phenomenon in
$2+1$-dimensions. We have derived the noncommutative (NC)
Chern-Simons action by integrating out the fermionic matter in the
one loop effective action in the long wavelength limit. This
result is valid to all orders in the NC parameter $\theta$.
Moreover the limit $\theta\rightarrow 0$ is smooth so that the NC
expressions reduce to their commutative versions and we recover
the bosonization in commutative spacetime.

Finally we return to the ambiguity \cite{sg,btc,wot} surrounding
the issue of duality in noncommutative Maxwell-Chern-Simons theory
and Self Dual theory and point out that the analysis of \cite{sg}
holds as far as the duality of the noncommutative versions of the
above two bosonic theories are concerned. However, if one wants to
tie up the bosonization of the noncommutative fermion theory as
well, (as is true in the commutative spacetime), the present work
shows that one needs to consider the noncommutative Chern-Simons
theory \cite{btc,wot}.

\appendix

\section{Some trace identities}

We consider a hermitian representation of the $2\times 2$ gamma
matrices satisfying $\gamma^{\mu}\gamma^{\nu}= g^{\mu\nu} + i
\epsilon^{\mu\nu\sigma}\gamma^{\sigma}$. This implies the
following trace identities in 2+1 dimensions:
\begin{eqnarray} \nonumber
{\rm tr\,}(\gamma^{\mu} \gamma^{\nu})&=& 2 g^{\mu \nu} \\
\nonumber
{\rm tr\,}(\gamma^{\mu} \gamma^{\nu} \gamma^{\sigma})&=&
2i\epsilon^{\mu\nu\sigma}  \\ \nonumber
{\rm tr\,}(\gamma^{\mu} \gamma^{\nu} \gamma^{\sigma}
\gamma^{\lambda}) &=& 2 \left(g^{\mu\nu} g^{\sigma\lambda} +
g^{\mu\lambda} g^{\nu\sigma} - g^{\mu\sigma} g^{\nu\lambda}\right)
\\ \nonumber
{\rm tr\,}(\gamma^{\mu} \gamma^{\nu} \gamma^{\sigma}
\gamma^{\lambda} \gamma^{\alpha}) &=& 2i \left(g^{\mu\nu}
\epsilon^{\sigma\lambda\alpha} + g^{\lambda\alpha}
\epsilon^{\mu\nu\sigma} +
g^{\sigma\lambda} \epsilon^{\mu\nu\alpha} - g^{\sigma\alpha} \epsilon^{\mu\nu\lambda}\right), \nonumber \\
{\rm tr\,}(\gamma^{\mu} \gamma^{\nu} \gamma^{\sigma}
\gamma^{\lambda} \gamma^{\alpha} \gamma^{\beta}) &=& 2
\left(g^{\mu\nu} g^{\sigma\lambda} g^{\alpha\beta} + g^{\mu\nu}
g^{\sigma\beta} g^{\lambda\alpha} - g^{\mu\nu} g^{\sigma\alpha}
g^{\lambda\beta} - g^{\alpha\beta} g^{\mu\sigma} g^{\nu\lambda}
\right. \nonumber \\
& + & \left. g^{\alpha\beta} g^{\mu\lambda} g^{\nu\sigma} -
g^{\lambda\alpha} g^{\mu\sigma} g^{\nu\beta} + g^{\lambda\alpha}
g^{\mu\beta} g^{\nu\sigma} + g^{\lambda\beta} g^{\mu\sigma}
g^{\nu\alpha} \right. \nonumber \\ \nonumber & - & \left.
g^{\lambda\beta} g^{\mu\alpha} g^{\nu\sigma} - g^{\mu \lambda}
g^{\nu \alpha} g^{\sigma \beta} - g^{\mu \alpha} g^{\nu \beta}
g^{\sigma \lambda} - g^{\mu \beta} g^{\nu \lambda} g^{\sigma
\alpha} \right. \\  \nonumber & + & \left. g^{\mu \lambda} g^{\nu
\beta} g^{\sigma \alpha} + g^{\mu \beta} g^{\nu \alpha} g^{\sigma
\lambda} + g^{\mu \alpha} g^{\nu \lambda} g^{\sigma \beta}
\right), \nonumber \\ \nonumber
\mathrm{tr} \, \left(\gamma^\rho \gamma^\tau \gamma^\mu \gamma^\nu
\gamma^\sigma \gamma^\lambda \gamma^\alpha \right) & = & 2 \, i \,
\left( g^{\rho \tau} g^{\mu \nu} \epsilon^{\sigma \lambda \alpha}
+ g^{\rho \tau} g^{\lambda \alpha} \epsilon^{\mu \nu \sigma} +
g^{\rho \tau} g^{\sigma \lambda} \epsilon^{\mu \nu \alpha} -
g^{\rho \tau} g^{\sigma \alpha} \epsilon^{\mu \nu \lambda} \right.
\\ \nonumber & + & \left. g^{\mu \nu} g^{\sigma \lambda}
\epsilon^{\rho \tau \alpha} + g^{\mu \nu} g^{\lambda \alpha}
\epsilon^{\rho \tau \sigma} - g^{\mu \nu} g^{\sigma \alpha}
\epsilon^{\rho \tau \lambda} - g^{\mu \sigma} g^{\nu \lambda}
\epsilon^{\rho \tau \alpha} \right. \\ \nonumber & + & \left.
g^{\mu \lambda} g^{\nu \sigma} \epsilon^{\rho \tau \alpha} -
g^{\lambda \alpha} g^{\mu \sigma} \epsilon^{\rho \tau \nu} +
g^{\lambda \alpha} g^{\nu \sigma} \epsilon^{\rho \tau \mu} +
g^{\mu \sigma} g^{\nu \alpha} \epsilon^{\rho \tau \lambda} \right.
\\ \nonumber
& - & \left. g^{\mu \alpha} g^{\nu \sigma} \epsilon^{\rho \tau \lambda} - g^{\mu \lambda} g^{\nu \alpha} \epsilon^{\rho \tau \sigma} - g^{\mu \alpha} g^{\sigma \lambda} \epsilon^{\rho \tau \nu} - g^{\nu \lambda} g^{\sigma \alpha} \epsilon^{\rho \tau \mu} \right. \\
& + & \left. g^{\mu \lambda} g^{\sigma \alpha} \epsilon^{\rho \tau
\nu} + g^{\nu \alpha} g^{\sigma \lambda} \epsilon^{\rho \tau \mu}
+ g^{\mu \alpha} g^{\nu \lambda} \epsilon^{\rho \tau \sigma}
\right).
\end{eqnarray}

\section{Some important integrals}

We also need the following forms of the integrals  \cite{ramallo}.
Note that the results presented below hold in the large mass limit
i.e., $m \rightarrow \infty$
\begin{eqnarray} \label{2aint}
\int \frac{d^3 q}{(2\pi)^3} \frac{1}{[(q + p_1)^2 - m^2 + i
\epsilon] [q^2 - m^2 + i \epsilon]} & = & \frac{i}{8 \pi m} \\
\label{3aint1} \int \frac{d^3 q}{(2\pi)^3} \frac{1}{[(q + p_1)^2 -
m^2 + i \epsilon] [q^2 - m^2 + i \epsilon][(q - p_2)^2 - m^2 + i
\epsilon]} & = & - \frac{1}{4} \left(\frac{i}{8 \pi m^3}\right) \\
\label{3aint2} \int \frac{d^3 q}{(2\pi)^3} \frac{q^2}{[(q + p_1)^2
- m^2 + i \epsilon] [q^2 - m^2 + i \epsilon][(q - p_2)^2 - m^2 + i
\epsilon]} & = & \frac{3}{4} \left(\frac{i}{8 \pi m}\right)
\end{eqnarray}

\end{document}